\documentclass[twocolumn, pra,showpacs,nofootinbib]{revtex4}
\usepackage{dcolumn,bm,graphicx}
\begin{document}
\preprint{UNR Mar 2001-\today }
\title{ Effects of confinement on permanent electric-dipole moment of Xe atoms in
liquid Xe }
\author{ Boris Ravaine}
\author{Andrei Derevianko}
\affiliation { Department of Physics, University of Nevada, Reno, Nevada 89557}

\date{\today}

\begin{abstract}
    Searches for permanent electric-dipole moments (EDM) of atoms provide
    important constraints on competing extensions to the standard model of
     elementary particles. Recently proposed experiment with liquid $^{129}$Xe
     [M.V. Romalis and M.P. Ledbetter, Phys.\ Rev.\ Lett.\ \textbf{87},
  067601 (2001)] may significantly improve present limits on the EDMs.
    To interpret experimental data in terms of CP-violating sources, one
    must relate measured atomic EDM to various model interactions
    via electronic-structure calculations.
    Here we study density dependence of atomic EDMs. The analysis
    is carried out in the framework of the cell model of the liquid coupled
    with relativistic atomic-structure calculations. We find
    that compared to an isolated atom, the EDM of an atom of liquid Xe
    is suppressed by about 40\%.
\end{abstract}


\maketitle


Most extensions of the standard model of elementary particles,
e.g., supersymmetry, naturally produce permanent electric dipole
moments (EDM) of atoms and molecules~\cite{KhrLam97} that are
comparable to or larger than present limits (see, e.g., a popular
review~\cite{ForPatBar03}). For example, the most accurate to
date determination of atomic EDM of $^{199}$Hg \cite{RomGriJac01}
sets limits on a number of important parameters: CP-violating QCD
vacuum angle, quark chromo-EDMs, semileptonic CP-violating
parameters, and restricts parameter space for certain extensions
to the standard model.
A substantial,  several orders of magnitude improvement
in sensitivity to all the enumerated sources of CP-violation  is anticipated
in an experiment proposed by~\citet{RomLed01}.
These authors propose to search for an EDM of a liquid sample of $^{129}$Xe.
Compared to the gas-phase experiment~\cite{RosChu01},
a drastically improved sensitivity of the liquid Xe experiment is mainly
due to the higher number densities of the liquid phase ($10^{22}\,
\mathrm{atoms}/\mathrm{cm}^3$).

The very use of the liquid phase raises questions about density-dependent
factors which can influence the outcome and interpretation of the experiment.
For example, an EDM experiment with a molecular liquid was proposed in
Ref. ~\cite{VarGorEzh82}. The authors found an
additional suppression of the EDM signal by a factor of a hundred due to
a reduced population of molecular rotational levels in
liquid. Although the experiment with  liquid Xe will be free from such an
effect, it is clear that the effects of the liquid phase on atomic
EDMs have to be investigated.

An EDM of an atom is related to a strength
of a CP-violating source via electronic-structure (enhancement or shielding) factors.
For an isolated Xe atom such factors
were computed previously: P,T-odd semileptonic
interactions were considered by \citet{Mar85} and
nuclear Schiff moment ---  by \citet{DzuFlaGin02}.
Here we employ a simple cell model
to study density dependence of the electronic-structure factors.
Technically, we extend the previous
atomic relativistic many-body calculations by
confining a Xe atom to a spherically-symmetric cavity.
In a non-polar liquid such as liquid Xe,
this cavity roughly approximates an averaged interaction
with the neighboring atoms. Imposing proper
boundary conditions at the cavity radius, first we solve
the Dirac-Hartree-Fock (DHF) equations and then
employ the relativistic random-phase approximation (RRPA) to
account for correlations. To the best of our knowledge,
here we report the first {\em ab initio} relativistic calculations
of properties of a liquid.
We find that compared to the EDM of an
isolated atom, the resulting EDM of an
atom of liquid Xe is suppressed by about 40\%.
Thus if the experiment with liquid Xe is carried out with
the anticipated sensitivity, we expect that the inferred constraints
on possible sources of CP-violation would be indeed
several orders of magnitude better than the present limits.



{\em Sources of atomic EDM --- }
The conventional atomic Hamiltonian $H_0$ among other symmetries is invariant with
respect to space-reflection (P) and time reversal (T). Therefore, on very general
grounds,
an expectation value of the electric dipole operator $\mathbf{D}=-\sum_{i}\mathbf{r}_i$
in a non-degenerate atomic state $| \Psi_0 \rangle $ vanishes.
The tiny CP-violating
interactions, here generically denoted as
$H_\mathrm{CP}=\sum_i h_\mathrm{CP}(\mathbf{r}_i)$,
break the symmetry of the atom and induce
a correction to the electronic state
$
|\tilde{\Psi}\rangle=|\Psi_{0}\rangle + |\delta\Psi\rangle \, .
$
To the lowest order
\begin{equation}
|\delta\Psi\rangle = \sum_k |\Psi_k\rangle \,
\frac{\langle \Psi_k| H_\mathrm{CP}| \Psi_0 \rangle }{E_0-E_k} \, ,
\label{Eq:deltaPsi}
\end{equation}
where $E_k$ and $|\Psi_{k}\rangle$ are eigenvalues and
eigenfunctions of $H_0$. Due to selection rules, the
$|\delta\Psi\rangle$ admixture has a
parity opposite to the one of the reference state $| \Psi_0 \rangle$.
Because of this opposite-parity admixture the atom acquires
a permanent EDM
\begin{equation}
\mathbf{d}=\langle\tilde{\Psi}|\mathbf{D}|\tilde{\Psi}\rangle =
2 \langle\Psi_{0}|\mathbf{D}|\delta\Psi\rangle \, .
\label{Eq:dExpectation}
\end{equation}

Now we specify particular forms of $H_{CP}$.
An analysis~\cite{KhrLam97} shows that
for diamagnetic atoms, such as Xe, the EDM predominantly arises due to
P,T-odd semi-leptonic interaction $H_\mathrm{TN}$ between electrons and nucleons
and also due to interaction $H_\mathrm{SM}$ of electrons
with the so-called nuclear Schiff moment~\cite{Sch63}. Smaller atomic EDM is
generated by intrinsic EDM of electrons and we will not consider
this mechanism here. Atomic units $|e|=\hbar=m_e=4\pi\varepsilon_0\equiv 1$ are used
throughout.


Explicitly, the effective P,T-odd semileptonic interaction Hamiltonian
may be represented as~\cite{Mar85}
\begin{equation}
h_\mathrm{TN}(\mathbf{r}_e)=\sqrt{2}G_\mathrm{F}\,C_\mathrm{TN}\,\bm{\sigma}_{N}\cdot\left(
i\gamma _{0}\gamma_{5}\,\bm{\sigma}\right) _{e}\rho_{N}\left(
\mathbf{r}_{e}\right)  \, .
\label{Eq:hTN}
\end{equation}
Here subscripts $e$ and $N$ distinguish between operators
acting in the space of electronic and nuclear coordinates respectively.
$C_\mathrm{TN}$ is a coupling constant to be determined from
an interpretation of EDM measurements and to be compared with
theoretical model-dependent predictions.
Due to averaging over nuclear degrees of freedom,
this interaction  depends on nuclear density
distribution $\rho_N(r)$. In the following,
we approximate $\rho_N(r)$ as a Fermi distribution $\rho_N(r)=\rho_0/(1+\exp[(r-c)/a])$
with
$c=5.6315$ fm and  $a=0.52$ fm.
Finally, $G_\mathrm{F}\approx 2.22254 \times 10^{-14}$ a.u. is the Fermi constant.

The interaction of an electron with the nuclear
Schiff moment $\mathbf{S}$
has the form~\cite{FlaGin02}
\begin{equation}
h_\mathrm{SM}(\mathbf{r}_e)=
\frac{3}{B_{4}}\,\rho_{N}\left( \mathbf{r}_e\right)
\left( \mathbf{r}_e\cdot\mathbf{S}\right) \, ,
\label{Eq:hSM}
\end{equation}
where $B_{4}=\int_{0}^{\infty}r^{4}\rho_{N}\left(  r\right) dr$ is
the fourth-order moment of the nuclear distribution.
The Schiff moment characterizes a difference between charge and
EDM distributions inside the nucleus.
It depends on a number of important
CP-violating parameters enumerated in the introduction.

Finally, we emphasize that
both  $h_\mathrm{TN}$ and $h_\mathrm{SM}$
are contact interactions. They occur when an electron penetrates the
nucleus. The electron speed at the nucleus is
approximately $\alpha Z c \simeq \frac{1}{2} c $
($Z=54$), i.e., a fully relativistic description
of electronic motion is important in this problem.

{\em Cell model of liquid xenon ---}
Here we employ a simple cell model (see \cite{Pat02} and references therein)
to estimate the effects of the environment on permanent
EDM of a given atom.
According to the cell model, we confine an
atom to a spherical cavity of radius
\begin{equation}
R_\mathrm{cav} = \left(\frac{3}{4\pi}\, \frac{1}{n}  \right)^{1/3} \, ,
\label{Eq:Rcav}
\end{equation}
$n$ being the number
density of the sample. For a density
of liquid Xe of 500 amagat~\cite{AmagatDef},
$R_\mathrm{cav} \simeq 4.9 $ bohr.
In non-relativistic calculations periodicity
requires that the normal component of the gradient of electronic
wave-function vanishes at the surface of the cell (see, e.g.,~\cite{StaBen91})
\begin{equation}
\frac{\partial \Psi}{\partial r} (R_\mathrm{cav})=0 \, .
\label{Eq:NRPeriodicBC}
\end{equation}

Before proceeding with a technical question of implementing
these boundary conditions in relativistic calculations, we
notice that the cell model implicitly incorporates
an average polarization interaction with the media. Indeed,
the  Hamiltonian of an atom placed in the liquid  in addition to
the conventional atomic Hamiltonian $H_0$ includes  interaction
of electrons with the rest of the atoms in the media. This
interaction is dominated by polarization potential. An important
point is that the {\em averaged} polarization interaction can be expressed as
$V_p=-1/2 (1-\epsilon^{-1}) R_\mathrm{cav}^{-1}$, where $\epsilon$
is the dielectric constant of the media~\cite{StaBen91}.
This interaction does not depend on electronic coordinate ---
it is just an additive constant which does not affect calculations of EDM.
Thus we may approximate the total Hamiltonian with the traditional
atomic Hamiltonian $H_0$.


Further, the spherical symmetry of the cell  allows us to
employ traditional methods of atomic structure.
The only modification is due to boundary conditions~(\ref{Eq:NRPeriodicBC}).
However, in relativistic calculations,
a special care should be taken when implementing this boundary condition.
Indeed, the Dirac bi-spinor may be represented as
\begin{equation}
 \varphi_{n \kappa m}(\mathbf{r})=\frac{1}{r}\left( \begin{array}{c} P_{n \kappa}(r)\,\,\,\Omega_{\kappa m}(\widehat{\mathbf{r}}) \\
i Q_{n \kappa}(r) \,\,\,\Omega_{-\kappa m}(\widehat{\mathbf{r}})
\end{array} \right) \, ,
\label{Eq:BiSpinor}
\end{equation}
where $P$ and $Q$ are the large and small radial components respectively
and $ \Omega$ is the spherical spinor. The angular quantum number
$\kappa=(l-j)\left(2j+1\right)$.
The nonrelativistic boundary condition~(\ref{Eq:NRPeriodicBC})
applied directly to the above
ansatz would lead to {\em two} separate constraints on  $P$ and $Q$.
This over-specifies boundary conditions and leads to the Klein
paradox.

A possible relativistic
generalization of the boundary condition~(\ref{Eq:NRPeriodicBC})
is 
\begin{equation}
\frac{d }{dr}\frac{P_{n \kappa}}{r} (R_\mathrm{cav})=
\frac{d }{dr}\frac{Q_{n \kappa}}{r} (R_\mathrm{cav}) \, .
\label{Eq:bagBCperiodic}
\end{equation}
Since in the non-relativistic limit the small component $Q$ vanishes,
this generalization subsumes Eq.~(\ref{Eq:NRPeriodicBC}).
Due to a semi-qualitative nature of our calculations,
here we have chosen to use simpler (MIT bag model) boundary condition
\begin{equation}
P_{n \kappa}(R_\mathrm{cav})=Q_{n \kappa}(R_\mathrm{cav}) \,.
\label{Eq:bagBC}
\end{equation}
Non-relativistically it corresponds to impenetrable
cavity surface. Compared to this condition,
the periodic boundary conditions~(\ref{Eq:bagBCperiodic}) are
``softer'', i.e., they modify the free-atom wavefunctions less significantly;
we expect that
our use of Eq.~(\ref{Eq:bagBC}) would somewhat overestimate the effects of
confinement in the liquid.

{\em Atom in a cavity: DHF and RRPA solutions ---}
To reiterate the discussion so far,  within the cell model,
the complex liquid-structure
problem is reduced to solving atomic many-body Dirac equation with
boundary conditions~(\ref{Eq:bagBC}). The atomic-structure
analysis is simplified by the fact that Xe is a closed-shell atom.
Below we self-consistently solve the DHF equations
inside the cavity.
Then we employ more sophisticated RRPA.

 At the DHF
level, the atomic wavefunction is represented by the Slater determinant
composed of occupied (core) orbitals $\varphi_a$. These orbitals are
determined from a set of DHF equations
\begin{equation}
\left( c (\bm{\alpha} \cdot \mathbf{p}) + \beta c^2 +
V_\mathrm{nuc} + V_\mathrm{DHF} \right) \varphi_a =
\varepsilon_a \varphi_a \, ,
\end{equation}
where $V_\mathrm{nuc}$ is a potential of the Coulomb interaction
with a finite-size nucleus of charge density $\rho_N(r)$
and $V_\mathrm{DHF}$ is non-local self-consistent DHF potential.
The DHF potential depends on all the core orbitals.
Similar equations may be written for (virtual) excited
orbitals $\varphi_m$.

We solved the DHF equations in the cavity using a B-spline
basis set technique by \citet{JohBluSap88}.
This technique is based on the Galerkin method:
the DHF equations are expressed in
terms of an extremum of an action integral $S_A$.
The boundary conditions are incorporated in the $S_A$ as well.
Further, the action
integral is expanded in terms of a finite set of basis functions (B-splines).
Minimization of such $S_A$  with respect to expansion coefficients
reduces solving integro-differential DHF equations to solving
symmetric generalized eigenvalue problem of linear algebra. The resulting set of
basis functions is finite and can be considered as numerically complete.
 In a typical calculation
we used a set of basis functions expanded over 100 B-splines.


Given a numerically complete set of DHF eigenfunctions $\{\varphi_i\}$,
the permanent atomic EDM, Eq.(\ref{Eq:dExpectation}), may be expressed as
\begin{equation}
\mathbf{d}^\mathrm{DHF}= 2 \sum_{m,a}\frac{\langle
                  \varphi_{a}|\mathbf{r}|\varphi_{m}\rangle\langle
                  \varphi_{m}|h_\mathrm{CP}|\varphi_{a}\rangle}
                  {\varepsilon_{m}-\varepsilon_{a}} \, ,
\label{Eq:dDHF}
\end{equation}
where $a$ runs over occupied and $m$ over virtual orbitals.
Here $h_\mathrm{CP}$ is either a semileptonic interaction, Eq.~(\ref{Eq:hTN}),
or an interaction with the nuclear Schiff moment, Eq.~(\ref{Eq:hSM}).
An additional peculiarity related to the Dirac equation is
an appearance of negative energy states ($\varepsilon_m < -m_e c^2$)
in the summation over intermediate states in Eq.~(\ref{Eq:dDHF}).
We have verified that these states introduce a completely
negligible correction to the computed EDMs.

To improve upon the DHF approximation, we have also
computed EDMs using RRPA method~\cite{FetWal71}.
This approximation
describes a dynamic linear response of
an atom to a perturbing one-body interaction (e.g., $H_\mathrm{CP}$).
The perturbation modifies core orbitals thus changing
the DHF potential. This modification of $V_\mathrm{DHF}$
in turn requires the orbitals to adjust self-consistently.
Such a readjustment process defines an infinite series of
many-body diagrams, shown, e.g., in Ref.~\cite{Mar85}.
The RRPA series can be summed  to all orders using iterative techniques
or solving DHF-like equations. We used an alternative method of solutions
based on the use of basis functions~\cite{Joh88}. As an input,
we used the DHF basis functions generated in the cavity (see discussion
above), i.e. the boundary conditions were satisfied automatically.
As a result of solving the RRPA equations, we have determined
a quasi-complete set of particle-hole excited states
and their energies. Then the EDMs are determined using
expressions similar to Eq. (\ref{Eq:deltaPsi})
and (\ref{Eq:dExpectation}).

{\em Discussion and conclusions ---}
First, we present the results of our calculations
for an isolated atom ($R_\mathrm{cav}=\infty$).
For the Schiff-moment-induced EDM, our results,
\begin{eqnarray*}
d^\mathrm{DHF}_\mathrm{SM} &=& 2.88
\,\left(\frac{S}{e\,\mathrm{fm}^{3}}\right ) \times 10^{-18} \,e\, \mathrm{cm} , \\
d^\mathrm{RRPA}_\mathrm{SM} &=& 3.78
\,\left(\frac{S}{e\,\mathrm{fm}^{3}}\right ) \times 10^{-18} \,e\, \mathrm{cm}
\, ,
\end{eqnarray*}
are in  agreement with the recent calculations by \citet{DzuFlaGin02}.
For the EDM induced by T,P--odd semileptonic interactions we obtain
\begin{eqnarray*}
d^\mathrm{DHF}_\mathrm{TN} &=& 8.44
 \times 10^{-13}\,\, C_\mathrm{TN}\, \sigma_{N} \, \mathrm{a.u.} \, , \\
d^\mathrm{RRPA}_\mathrm{TN} &=& 10.7
 \times 10^{-13}\,\, C_\mathrm{TN}\, \sigma_{N} \, \mathrm{a.u.}.
\end{eqnarray*}
These values are to be compared with the results by \citet{Mar85},
$d^\mathrm{DHF}_\mathrm{TN}= 7.764$ and $d^\mathrm{RRPA}_\mathrm{TN}= 9.808$
in the same units. The reason for the 10\% difference between our results
and those from Ref.~\cite{Mar85} is not clear.

Before presenting results for finite cavity radii, let us consider
individual contributions to EDM from various shells of Xe atom.
These contributions for
the Schiff-moment-induced EDM of an isolated atom
are listed in Table~\ref{table:Sbreak}. A similar table, but
for the EDM arising from semileptonic
interactions is given in Ref.~\cite{Mar85}.
From these tables  we observe that the dominant contribution to EDMs comes
from the outer $n=5$ shell. Thus we anticipate that a noticeable density
dependence should occur when $R_\mathrm{cav}$ becomes comparable to the size of
external $n=5$ shell.  We also
notice that the contribution from the outer shell
is relatively more important in RRPA calculations than at the DHF level,
i.e., the RRPA results should exhibit stronger density dependence.

\begin{center}
\begin{table}[h]
\begin{tabular}{ldd}
\hline\hline
\multicolumn{1}{c}{} &
\multicolumn{1}{c}{DHF } &
\multicolumn{1}{c}{RRPA}    \\
\hline
$n=1$ & 0.039 & 0.039 \\
$n=2$ & 0.091 & 0.092 \\
$n=3$ & 0.20  & 0.21 \\
$n=4$ & 0.52  & 0.64 \\
$n=5$ & 2.0   & 2.8 \\
\hline
Total & 2.88  & 3.78  \\
\hline\hline
\end{tabular}
\caption{Individual contributions from various
shells to the EDM of a free $^{129}$Xe atom in the DHF and RRPA methods.
The EDM is induced by the nuclear Schiff
moment and it is given in units of
$ S/(e \,\mathrm{fm}^{3} ) \times 10^{-18} \,e\, \mathrm{cm}$.
\label{table:Sbreak}
}
\end{table}
\end{center}

\begin{figure}[h]
\begin{center}
\includegraphics*[scale=0.6]{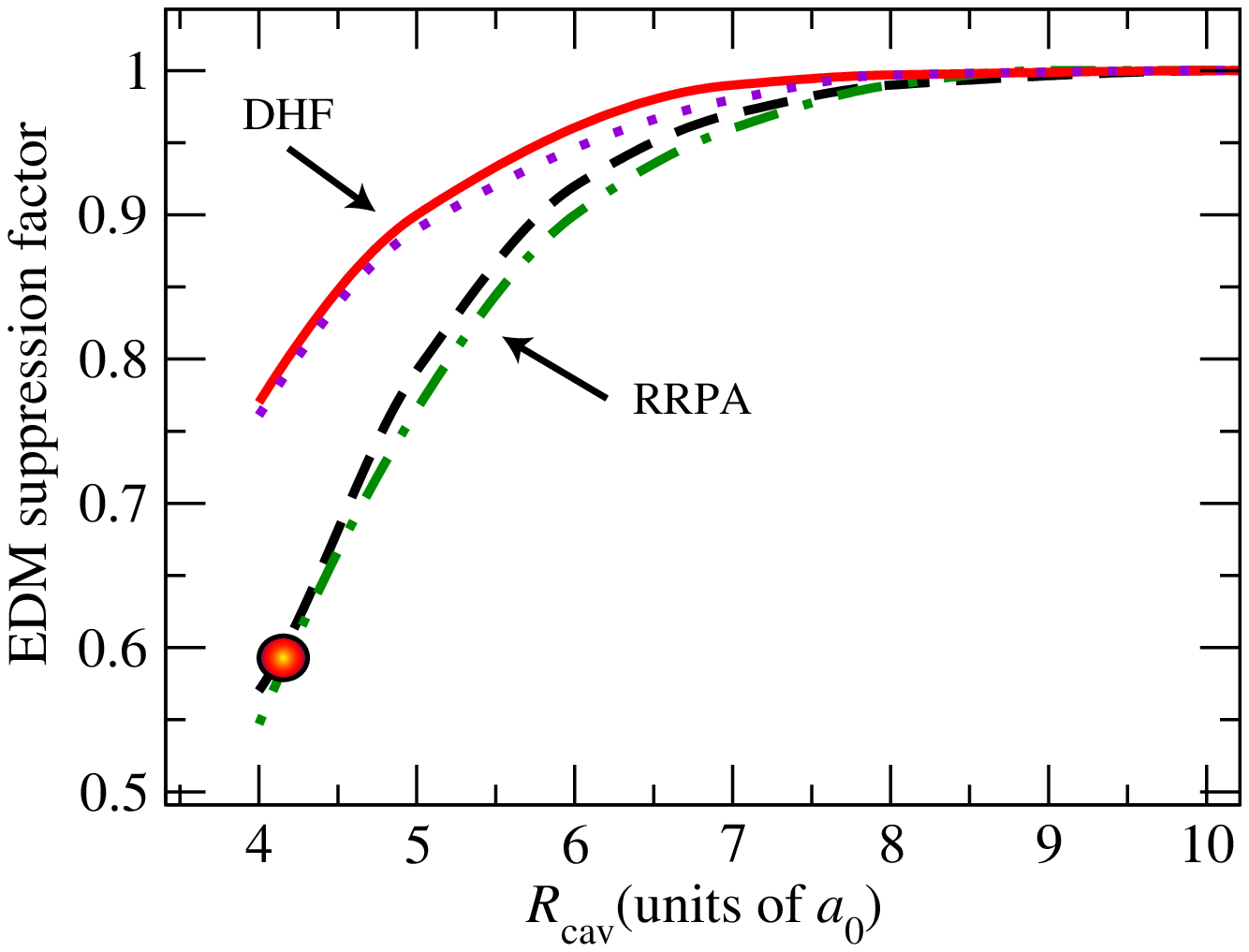
}
\caption{ The ratios of atomic EDMs for
the confined and isolated atoms (suppression factor)
as a function of cavity radius. The upper and lower
sets of two curves are obtained with the DHF and RRPA methods
respectively. EDMs induced by P,T--odd semileptonic interactions
are shown as solid and dashed lines, while EDMs due to the Schiff moment ---
as dotted and dashed-dotted lines. Heavy dot marks our final results
for liquid Xe.
 \label{Fig:EDM}} \end{center}
\end{figure}

These qualitative conclusions for a confined atom
are supported by our numerical results, presented
in Fig.~\ref{Fig:EDM}.
 Here we plot the ratios of atomic EDMs for
the confined and isolated atoms as a function of $R_\mathrm{cav}$.
The EDMs become smaller as the density
increases, $n \propto R_\mathrm{cav}^{-3}$.
At the density of liquid Xe, $R_\mathrm{cav} \approx 4.9$ bohr ,
the more accurate RRPA results
show a 25\% suppression of the atomic EDM due to confinement.
Overall there is a noticeable density-dependence of atomic EDM.
We expect the EDM signal (if found) to be broadened. The
relevant characteristic width of the signal can be simply
estimated from our Fig.~\ref{Fig:EDM} from the mean density fluctuations.

>From Fig.~\ref{Fig:EDM} we notice that
both semileptonic-- and Schiff--moment--induced EDMs
scale with $R_\mathrm{cav}$ in a  similar fashion.
This similarity can be explained from the following
arguments. The values of  CP-violating matrix elements,
Eq.(\ref{Eq:hTN}) and Eq.(\ref{Eq:hSM}),
are accumulated inside the nucleus. Non-relativistically,
as $r \rightarrow 0$
the wavefunctions scale
as $\varphi_{nlm}(\mathbf{r}) \approx N_{nl}(R_\mathrm{cav}) \times r^{l} Y_{lm}( \hat{\mathbf{r}} )$,
where $N_{nl}$ are normalization factors.
Therefore the dominant contribution to the EDM, Eq.(\ref{Eq:dDHF})
arises from mixing of $s$ and $p$ states.  By factorizing the matrix element
of $h_\mathrm{CP}$ as $ \langle \varphi_{ns}|h_\mathrm{CP}|\varphi_{n'p}\rangle
 \approx  N_{ns}(R_\mathrm{cav}) N_{n'p}(R_\mathrm{cav}) \times
  \langle s |h_\mathrm{CP}| p \rangle $ we see that the
$R_\mathrm{cav}$--independent factor $\langle s |h_\mathrm{CP}| p \rangle $
can be pulled out of the summation
over atomic orbitals in Eq.(\ref{Eq:dDHF}).
Thus, both semileptonic-- and Schiff--moment--induced EDMs exhibit
approximately
the same scaling with the cavity radius.
A correction to this ``similarity scaling law''
may arise, for example, due to different selection rules
involved for the two EDM operators.


It is worth emphasizing the semi-qualitative nature
of our calculations. The analysis can be improved by
employing  more realistic models of liquid environment.
Even within the cell model  we could further refine our analysis.
 A dense liquid may be considered as a solid with vacancies,
i.e., the decrease of the average bond length is negligible,
rather the nearest-neighbor occupation numbers are decreased compared to
solid. Xe condenses into face-centered cubic structure. The first
nearest-neighbor shell contains twelve atoms (partially justifying
the spherical symmetry of the elementary cell).
The density of the solid Xe is $3.54$ g/cm$^3$, implying
the half-radius of this shell of 4.2 bohr, somewhat smaller than
$R_\mathrm{cav} \approx 4.9$ bohr for liquid Xe. As follows from
Fig.~\ref{Fig:EDM},
this difference leads to more pronounced suppression of the atomic
EDM by 40\% per cent.

To reiterate, our work was motivated by anticipated significant improvements in
sensitivity  to atomic EDMs in experiments with liquid $^{129}$Xe~\cite{RomLed01}.
Here we investigated confining effects of the environment
on the EDM of Xe atom.
We carried out the analysis in the framework
of the cell model coupled with relativistic atomic-structure
calculations. We found that compared to an isolated atom,
the EDM of an atom of liquid Xe is reduced by about 40\%.
Thus if the experiment with liquid Xe is carried out with
the anticipated sensitivity, we expect that the inferred constraints
on possible sources of CP-violation would be indeed
several orders of magnitude better than the present limits.


We would like to thank M. Romalis for discussions.
This work was supported in part by the National Science Foundation.

\bibliography{EDM,general,exact,rpa,XeEDMnotes}

\begin{thebibliography}{16}
\expandafter\ifx\csname natexlab\endcsname\relax\def\natexlab#1{#1}\fi
\expandafter\ifx\csname bibnamefont\endcsname\relax
  \def\bibnamefont#1{#1}\fi
\expandafter\ifx\csname bibfnamefont\endcsname\relax
  \def\bibfnamefont#1{#1}\fi
\expandafter\ifx\csname citenamefont\endcsname\relax
  \def\citenamefont#1{#1}\fi
\expandafter\ifx\csname url\endcsname\relax
  \def\url#1{\texttt{#1}}\fi
\expandafter\ifx\csname urlprefix\endcsname\relax\def\urlprefix{URL }\fi
\providecommand{\bibinfo}[2]{#2}
\providecommand{\eprint}[2][]{\url{#2}}

\bibitem[{\citenamefont{Khriplovich and Lamoreaux}(1997)}]{KhrLam97}
\bibinfo{author}{\bibfnamefont{I.~B.} \bibnamefont{Khriplovich}}
  \bibnamefont{and} \bibinfo{author}{\bibfnamefont{S.~K.}
  \bibnamefont{Lamoreaux}}, \emph{\bibinfo{title}{CP violation without
  strangeness. Electric dipole moments of particles, atoms, and molecules.}}
  (\bibinfo{publisher}{Springer}, \bibinfo{address}{Berlin},
  \bibinfo{year}{1997}).

\bibitem[{\citenamefont{Fortson et~al.}(2003)\citenamefont{Fortson, Sandars,
  and Barr}}]{ForPatBar03}
\bibinfo{author}{\bibfnamefont{E.~N.} \bibnamefont{Fortson}},
  \bibinfo{author}{\bibfnamefont{P.}~\bibnamefont{Sandars}}, \bibnamefont{and}
  \bibinfo{author}{\bibfnamefont{S.}~\bibnamefont{Barr}},
  \bibinfo{journal}{Physics Today} \textbf{\bibinfo{volume}{56(6)}},
  \bibinfo{pages}{33} (\bibinfo{year}{2003}).

\bibitem[{\citenamefont{Romalis et~al.}(2001)\citenamefont{Romalis, Griffith,
  Jacobs, and Fortson}}]{RomGriJac01}
\bibinfo{author}{\bibfnamefont{M.~V.} \bibnamefont{Romalis}},
  \bibinfo{author}{\bibfnamefont{W.~C.} \bibnamefont{Griffith}},
  \bibinfo{author}{\bibfnamefont{J.~P.} \bibnamefont{Jacobs}},
  \bibnamefont{and} \bibinfo{author}{\bibfnamefont{E.~N.}
  \bibnamefont{Fortson}}, \bibinfo{journal}{Phys. Rev. Lett.}
  \textbf{\bibinfo{volume}{86}}, \bibinfo{pages}{2505} (\bibinfo{year}{2001}).

\bibitem[{\citenamefont{Romalis and Ledbetter}(2001)}]{RomLed01}
\bibinfo{author}{\bibfnamefont{M.~V.} \bibnamefont{Romalis}} \bibnamefont{and}
  \bibinfo{author}{\bibfnamefont{M.~P.} \bibnamefont{Ledbetter}},
  \bibinfo{journal}{Phys.\ Rev.\ Lett.} \textbf{\bibinfo{volume}{87}},
  \bibinfo{pages}{067601} (\bibinfo{year}{2001}).

\bibitem[{\citenamefont{Rosenberry and Chupp}(2001)}]{RosChu01}
\bibinfo{author}{\bibfnamefont{M.~A.} \bibnamefont{Rosenberry}}
  \bibnamefont{and} \bibinfo{author}{\bibfnamefont{T.~E.} \bibnamefont{Chupp}},
  \bibinfo{journal}{Phys. Rev. Lett.} \textbf{\bibinfo{volume}{86}},
  \bibinfo{pages}{22} (\bibinfo{year}{2001}).

\bibitem[{\citenamefont{Varentsov et~al.}(1982)\citenamefont{Varentsov,
  Gorshkov, Ezhov, Kozlov, Labzovskii, and Fomichev}}]{VarGorEzh82}
\bibinfo{author}{\bibfnamefont{V.~L.} \bibnamefont{Varentsov}},
  \bibinfo{author}{\bibfnamefont{V.~G.} \bibnamefont{Gorshkov}},
  \bibinfo{author}{\bibfnamefont{V.~F.} \bibnamefont{Ezhov}},
  \bibinfo{author}{\bibfnamefont{M.~G.} \bibnamefont{Kozlov}},
  \bibinfo{author}{\bibfnamefont{L.~N.} \bibnamefont{Labzovskii}},
  \bibnamefont{and} \bibinfo{author}{\bibfnamefont{V.~N.}
  \bibnamefont{Fomichev}}, \bibinfo{journal}{Pis'ma Zh.\ Eksp.\ Teor.\ Fiz.}
  \textbf{\bibinfo{volume}{36}}, \bibinfo{pages}{141} (\bibinfo{year}{1982}),
  \bibinfo{note}{[JETP Lett. {\bf 36} 175 (1982)]}.

\bibitem[{\citenamefont{M{\aa}rtensson-Pendrill}(1985)}]{Mar85}
\bibinfo{author}{\bibfnamefont{A.}~\bibnamefont{M{\aa}rtensson-Pendrill}},
  \bibinfo{journal}{Phys. Rev. Lett.} \textbf{\bibinfo{volume}{54}},
  \bibinfo{pages}{1153} (\bibinfo{year}{1985}).

\bibitem[{\citenamefont{Dzuba et~al.}(2002)\citenamefont{Dzuba, Flambaum,
  Ginges, and Kozlov}}]{DzuFlaGin02}
\bibinfo{author}{\bibfnamefont{V.}~\bibnamefont{Dzuba}},
  \bibinfo{author}{\bibfnamefont{V.}~\bibnamefont{Flambaum}},
  \bibinfo{author}{\bibfnamefont{J.}~\bibnamefont{Ginges}}, \bibnamefont{and}
  \bibinfo{author}{\bibfnamefont{M.}~\bibnamefont{Kozlov}},
  \bibinfo{journal}{Phys. Rev. A} \textbf{\bibinfo{volume}{66}},
  \bibinfo{pages}{012111/1} (\bibinfo{year}{2002}).

\bibitem[{\citenamefont{Schiff}(1963)}]{Sch63}
\bibinfo{author}{\bibfnamefont{L.~I.} \bibnamefont{Schiff}},
  \bibinfo{journal}{Phys.\ Rev.} \textbf{\bibinfo{volume}{132}},
  \bibinfo{pages}{2194} (\bibinfo{year}{1963}).

\bibitem[{\citenamefont{Flambaum and Ginges}(2002)}]{FlaGin02}
\bibinfo{author}{\bibfnamefont{V.}~\bibnamefont{Flambaum}} \bibnamefont{and}
  \bibinfo{author}{\bibfnamefont{J.}~\bibnamefont{Ginges}},
  \bibinfo{journal}{Phys. Rev. A} \textbf{\bibinfo{volume}{65}},
  \bibinfo{pages}{032113} (\bibinfo{year}{2002}).

\bibitem[{\citenamefont{Patil}(2002)}]{Pat02}
\bibinfo{author}{\bibfnamefont{S.~H.} \bibnamefont{Patil}},
  \bibinfo{journal}{J. Phys. B} \textbf{\bibinfo{volume}{35}},
  \bibinfo{pages}{255} (\bibinfo{year}{2002}).

\bibitem[{Ama()}]{AmagatDef}
\bibinfo{note}{Amagat density unit is equal to 44.615 moles per cubic meter
  (mol/m$^3$)}.

\bibitem[{\citenamefont{Stampfli and Bennemann}(1991)}]{StaBen91}
\bibinfo{author}{\bibfnamefont{P.}~\bibnamefont{Stampfli}} \bibnamefont{and}
  \bibinfo{author}{\bibfnamefont{K.~H.} \bibnamefont{Bennemann}},
  \bibinfo{journal}{Phys. Rev. A} \textbf{\bibinfo{volume}{44}},
  \bibinfo{pages}{8210} (\bibinfo{year}{1991}).

\bibitem[{\citenamefont{Johnson et~al.}(1988)\citenamefont{Johnson, Blundell,
  and Sapirstein}}]{JohBluSap88}
\bibinfo{author}{\bibfnamefont{W.~R.} \bibnamefont{Johnson}},
  \bibinfo{author}{\bibfnamefont{S.~A.} \bibnamefont{Blundell}},
  \bibnamefont{and}
  \bibinfo{author}{\bibfnamefont{J.}~\bibnamefont{Sapirstein}},
  \bibinfo{journal}{Phys.\ Rev.\ A} \textbf{\bibinfo{volume}{37}},
  \bibinfo{pages}{307} (\bibinfo{year}{1988}).

\bibitem[{\citenamefont{Fetter and Walecka}(1971)}]{FetWal71}
\bibinfo{author}{\bibfnamefont{A.~L.} \bibnamefont{Fetter}} \bibnamefont{and}
  \bibinfo{author}{\bibfnamefont{J.~D.} \bibnamefont{Walecka}},
  \emph{\bibinfo{title}{Quantum Theory of Many-particle Systems}}
  (\bibinfo{publisher}{McGraw-Hill}, \bibinfo{year}{1971}).

\bibitem[{\citenamefont{Johnson}(1988)}]{Joh88}
\bibinfo{author}{\bibfnamefont{W.~R.} \bibnamefont{Johnson}},
  \bibinfo{journal}{Adv. At. Mol. Phys.} \textbf{\bibinfo{volume}{25}},
  \bibinfo{pages}{375} (\bibinfo{year}{1988}).

\end{thebibliography}

\end{document}